\documentclass[conference]{IEEEtran}
\IEEEoverridecommandlockouts
\usepackage{cite}
\usepackage{amsmath,amssymb,amsfonts}
\usepackage{algorithmic}
\usepackage{graphicx}
\usepackage{textcomp}
\usepackage{xcolor}
\usepackage[linesnumbered,ruled,vlined]{algorithm2e}
\usepackage[nolist,nohyperlinks]{acronym} 

\usepackage{float}
\usepackage{url}
\usepackage{graphics}
\usepackage{epstopdf}
\usepackage[linesnumbered,ruled,vlined]{algorithm2e}
\usepackage{algorithmic}
\usepackage{subcaption}

\def\BibTeX{{\rm B\kern-.05em{\sc i\kern-.025em b}\kern-.08em
    T\kern-.1667em\lower.7ex\hbox{E}\kern-.125emX}}
    
\definecolor{SolutionColor}{rgb}{0.9,0.9,1}

\def\gil#1{{\color{black}#1}}

\begin{document}

\begin{acronym}
\acro{STI}{short-term imperfection}
\acro{RIS}{Reconfigurable intellgent surface}
\acro{NMSE}{normalized mean square error}
\acro{HOSVD}{higher order singular value decomposition}
\end{acronym}

\title{Tensor-Based Channel Estimation for \ac{RIS}-Assisted Networks Operating Under Imperfections\\
}

\author{\IEEEauthorblockN{Paulo R. B. Gomes}
\IEEEauthorblockA{\textit{Wireless Telecom Research Group} \\
\textit{Federal University of Ceará}\\
Fortaleza, Brazil\\
paulo@gtel.ufc.br}
\and
\IEEEauthorblockN{Gilderlan T. de Araújo}
\IEEEauthorblockA{\textit{Wireless Telecom Research Group} \\
\textit{Federal University of Ceará}\\
Fortaleza, Brazil\\
gilderlan@gtel.ufc.br}
\and
\IEEEauthorblockN{Bruno Sokal}
\IEEEauthorblockA{\textit{Wireless Telecom Research Group} \\
\textit{Federal University of Ceará}\\
Fortaleza, Brazil\\
brunosokal@gtel.ufc.br}
\and
\IEEEauthorblockN{André L. F. de Almeida}
\IEEEauthorblockA{\textit{Wireless Telecom Research Group} \\
\textit{Federal University of Ceará}\\
Fortaleza, Brazil\\
andre@gtel.ufc.br}
\and
\IEEEauthorblockN{Behrooz Makki}
\IEEEauthorblockA{\textit{\textcolor{black}{Ericsson Research}}\\
Göteborg, Sweden \\
behrooz.makki@ericsson.com}
\and
\IEEEauthorblockN{Gábor Fodor}
\IEEEauthorblockA{\textit{\textcolor{black}{Ericsson Research and KTH Royal Institute of Technology}}\\ 
Stockholm, Sweden\\
gabor.fodor@ericsson.com}
}

\maketitle

\begin{abstract}
\textcolor{black}{Reconfigurable intelligent surface (RIS) is a candidate technology for future wireless networks.}
\textcolor{black}{It enables} \textcolor{black}{to shape the}
wireless environment to reach massive connectivity and enhanced data rate. The promising gains of \ac{RIS}-assisted networks \textcolor{black}{are, however,} \textcolor{black}{strongly depends} on the accuracy of the channel state information. 
\textcolor{black}{Due} to the passive nature of the \ac{RIS} elements, channel estimation \textcolor{black}{may become challenging}. This becomes most evident when physical \textcolor{black}{imperfections} or \textcolor{black}{electronic impairments} affect the \ac{RIS} due to its exposition to different environmental \textcolor{black}{effects} or caused by hardware limitations from the circuitry. \textcolor{black}{In this paper, we propose an efficient and low-complexity tensor-based channel estimation approach in \ac{RIS}-assisted networks taking different imperfections into account.} By assuming a short-term model \textcolor{black}{in which} 
the \ac{RIS} imperfections behavior, modeled as unknown amplitude and phase shifts deviations, \textcolor{black}{is}
non-static with respect to the channel coherence time, we formulate a closed-form higher order singular value decomposition based 
algorithm for the joint estimation of the involved channels and the unknown impairments. Furthermore, the identifiability and computational complexity of the proposed algorithm are analyzed, \textcolor{black}{and we study the effect of different imperfections on the channel estimation quality.} Simulation results demonstrate the effectiveness of \textcolor{black}{our} proposed tensor-based algorithm in terms of \textcolor{black}{the} estimation accuracy and computational complexity compared to competing tensor-based iterative alternating \textcolor{black}{solutions.} 
\end{abstract}
\begin{IEEEkeywords}
Reconfigurable intelligent surface, channel estimation, imperfections detection, hardware, MIMO, HOSVD. 
\end{IEEEkeywords}

\section{Introduction}

Wireless communications have become a necessity in our daily lives providing significant benefits to our modern society. Its popularization occurs thanks to frequent efforts and proliferation of technological innovations in order to \textcolor{black}{improve} 
system capacity and diverse quality of service requirements for different applications. However, \textcolor{black}{such achievement has led to} \textcolor{black}{the} massive increment in the number of connected devices as well as their rate \textcolor{black}{requirements} \cite{SCisco,RCITU}.
\textcolor{black}{As the network size and capabilities increase, energy consumption and implementation costs become challenging} \cite{SZhang2017}. \textcolor{black}{Therefore, to guarantee green sustainable wireless networks, one needs to carefully take the energy consumption and the hardware cost into account} \cite{QWu}.  

\textcolor{black}{Thanks} to the recent development of meta-materials, \textcolor{black}{reconfigurable intelligent surface (RIS)} \textcolor{black}{has} emerged as a potential technology applicable in future wireless networks through the novel \textit{smart and programming environment} paradigm. \textcolor{black}{Unlike conventional networks, \ac{RIS}-assisted networks enable the system to} \textcolor{black}{shape the} wireless environment \textcolor{black}{to be} suitable for \textcolor{black}{communication} \cite{DiRenzo2019}. \textcolor{black}{With a well-configured \ac{RIS}, the propagation conditions can be improved by controlling the scattering characteristics to create passive and active beamforming at the \ac{RIS} and the wireless transceivers to achieve high beamforming gains \cite{BehroozSug}.} In addition, a passive \ac{RIS} does not require active radio frequency 
chains for signal transmission, reception and processing \textcolor{black}{since}
it simply relies on passive signal reflection, which makes it a power-efficient, cost-effective, and \textcolor{black}{low-complexity} technology \cite{EBasar2019,QWuRZhang20192}.

\begin{figure}[!t]
	\centering
	\includegraphics[width=0.45\textwidth]{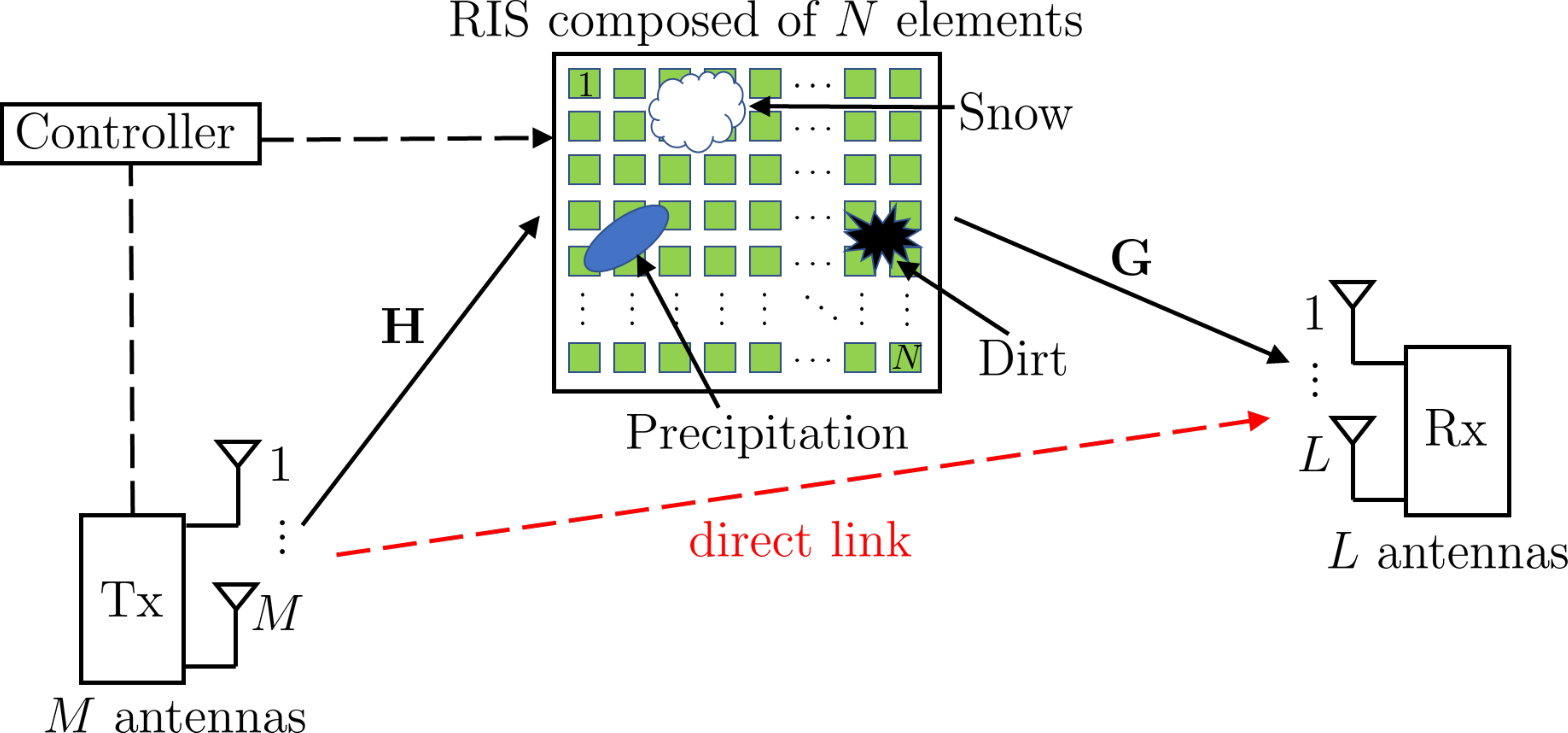}
	\caption{An illustration of a \ac{RIS}-assisted network operating under imperfections.}
	\label{figure1}
\end{figure}

Despite \textcolor{black}{its potential benefits,} the performance gains achieved in \ac{RIS}-assisted networks \textcolor{black}{are} strongly dependent on the quality of the channel state information (CSI), \textcolor{black}{among others.} \textcolor{black}{Also,} cost reduction may affect the hardware capability and, consequently, its performance. Therefore, to have a realistic view on the performance and usefulness of \ac{RIS}, one needs to carefully take the hardware impairments as well as the environmental effects, such as water precipitations, flying debris, air particles, snowflakes, ice stones, dry/damp sand particles and dirt into account, as shown in Fig. \ref{figure1}. 

In practice, 
such blocking objects as well as the hardware impairments induce unwanted attenuation and phase shifts on the reflected signals by the \ac{RIS} introducing time-varying distortions in the received signal, which directly affect the channel estimation accuracy and consequently the network performance. In this sense, typical channel estimation methods 
may not be able to deal with different imperfections and may fail to properly estimate the channel. Therefore, it is necessary to continuously monitor the channel and compensate for imperfections in order to maintain robust network operation. 

The effects of hardware impairments and environmental imperfections have been rarely studied in the literature. \textcolor{black}{For instance,} \cite{imp1,imp4,imp5}
\textcolor{black}{consider} \ac{RIS} operating under finite resolution of the phase shifts or phase estimation errors from imperfect channel estimation. More completely, \textcolor{black}{\cite{mainblock} and \cite{BiLi02}} consider different environmental effects on the \ac{RIS} and propose methods to jointly estimate the channel and array blockage parameters in millimeter wave (mmWave) \textcolor{black}{\ac{RIS}-assisted} systems. 

In this paper, we propose an efficient and low-complexity tensor-based algorithm for the joint estimation of the involved communication channels and imperfections in \ac{RIS}-assisted networks. First, we show that the received signal under the short-term imperfection (STI) model in which the RIS imperfections behavior \textcolor{black}{is} 
non-static with respect to the channel coherence time can be recast as a tensor following a quadrilinear parallel factor (PARAFAC) model. Exploiting the multi-linear structure of this model, we derive a closed-form higher order singular value decomposition (HOSVD)-based algorithm to solve the joint channel and \ac{RIS} imperfections estimation problem. Moreover, we also study the identifiability of the proposed estimator, discuss its computational complexity and investigate the effect of imperfections on the network performance. Compared to the typical channel estimation methods such as \cite{GilJournal}, \textcolor{black}{the key features of the proposed HOSVD-STI algorithm are its ability to properly estimate the channel and its robustness to different kinds of}
real-world imperfections at the \ac{RIS} with low computational complexity, since no iteration is required.

\textcolor{black}{\textbf{Notations.}} We use the following notations and properties. Scalars are denoted by lower-case letters ($a, \ldots$), column vectors by bold lower-case letters ($\mathbf{a}, \ldots$), matrices by bold upper-case letters ($\mathbf{A}, \ldots$) and tensors are represented by upper-case calligraphic letters ($\mathcal{A}, \ldots$). \textcolor{black}{Then,} $\mathbf{A}^{\text{T}}$ and $\mathbf{A}^{\dag}$ stand for the transpose and Moore-Penrose pseudo-inverse of $\mathbf{A}$, respectively. The operator $\text{vec}(\cdot)$ vectorizes its matrix argument by stacking its columns on top of each other, while $\text{vecd}(\cdot)$ forms a vector out of the diagonal of its matrix argument. \textcolor{black}{Also,} $\| \cdot \|_{\text{F}}$ represents the Frobenius norm \textcolor{black}{of a matrix or a tensor}, \gil{while $\times_n$ denotes the $n$-mode product}. \textcolor{black}{Moreover,} $\mathbf{I}_{M}$ is the $M \times M$ identity matrix. The operator $\mathbf{D}_{i}\left(\mathbf{A}\right)$ forms a diagonal matrix from the $i$-th row of $\mathbf{A}$, while the operator $\text{diag}(\mathbf{a})$ forms a diagonal matrix from $\mathbf{a}$. The Kronecker, Hadamard and the outer product operators are denoted by $\otimes$, $\odot$ and $\circ$, respectively. The Khatri-Rao product between two matrices is defined as
\begin{equation}
\mathbf{A} \diamond \mathbf{B} = \left[\mathbf{D}_{1}\left(\mathbf{A}\right)\mathbf{B}^{\text{T}}, \ldots, \mathbf{D}_{Q}\left(\mathbf{A}\right)\mathbf{B}^{\text{T}}\right]^{\text{T}},
\label{pp2}
\end{equation}
where $\mathbf{A} = \left[\mathbf{a}_{1}, \ldots, \mathbf{a}_{Q}\right] \in \mathbb{C}^{I \times Q}$ and $\mathbf{B} = \left[\mathbf{b}_{1}, \ldots, \mathbf{b}_{Q}\right] \in \mathbb{C}^{J \times Q}$.

In this paper, the following property of the Kronecker product will be used
\textcolor{black}{
\begin{eqnarray}
\mathbf{a} \otimes \mathbf{b} \otimes \mathbf{c} &=& \text{vec}\left(\mathbf{c} \circ \mathbf{b} \circ \mathbf{a}\right), \space \forall \space \mathbf{a}, \mathbf{b}, \mathbf{c}. \label{pp4}
\end{eqnarray}
}
\indent Furthermore, the definitions and operations involving tensors are in accordance with references \cite{Kolda} and \cite{hosvd}. 

\section{System Model}

\begin{figure}[!t]
	\centering
	\includegraphics[width=0.37\textwidth]{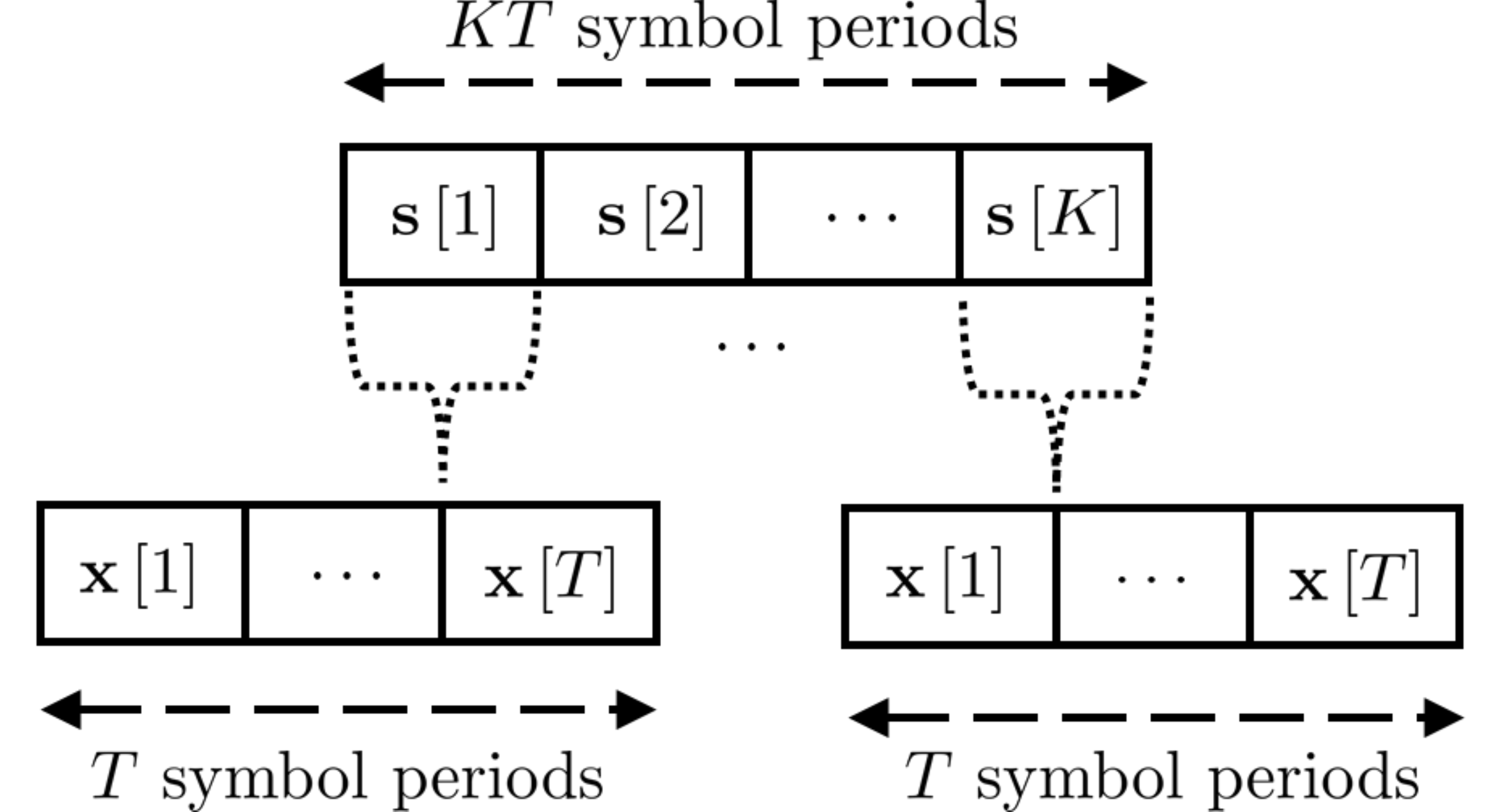}
	\caption{Structured pilot pattern in the time domain.}
	\label{protocol}
\end{figure}
\begin{figure}[!t]
   \includegraphics[width=1\linewidth]{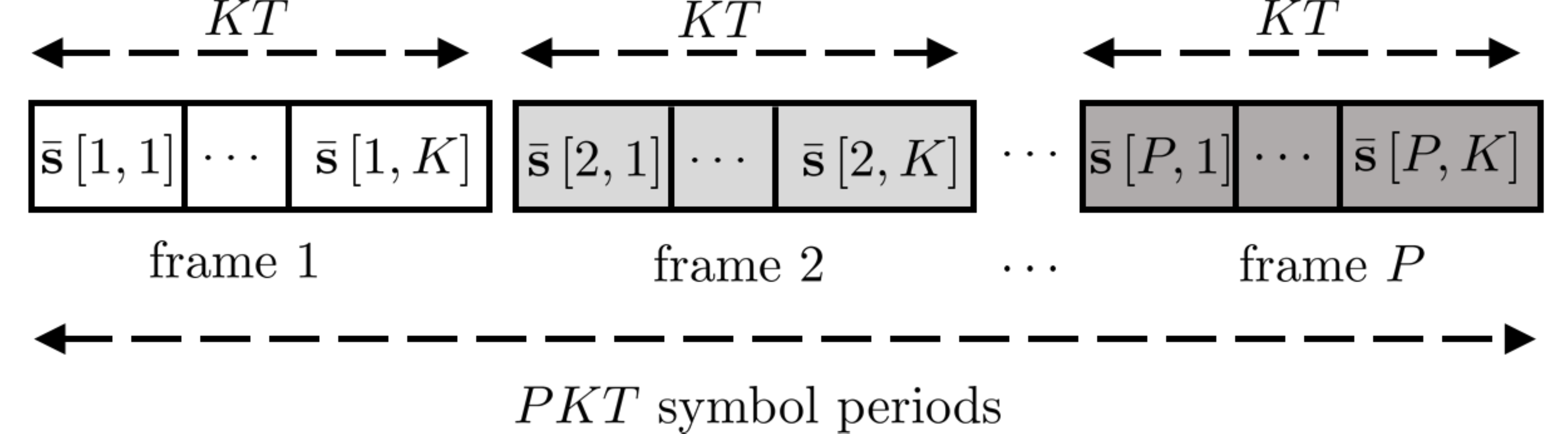}
   \caption{An illustration of the \textcolor{black}{STI model with time-varying \ac{RIS} fluctuations during the channel coherence time.}} 
   \label{sti}
\end{figure}
We consider a single-user narrowband \ac{RIS}-assisted MIMO network in which the transmitter (Tx) and the receiver (Rx) are equipped with arrays composed of $M$ and $L$ antennas, respectively, and the RIS has $N$ passive reflecting elements, \gil{as illustrated in \textcolor{black}{Fig.} \ref{figure1}}. 
The direct Tx-Rx link is assumed to be too weak or unavailable due to unfavorable propagation conditions. \gil{We assume a block-fading channel and the channel estimation occurs after the Tx sends pilot symbols which are reflected by the RIS using a known activation pattern and the receiver collects $p = 1, \dots, P$ frames.} Each frame is composed by $K$ blocks, $k = 1, \dots, K$, \textcolor{black}{each of} \textcolor{black}{which} contains $T$, $t = 1, \dots, T$ symbol periods. \textcolor{black}{Figure \ref{protocol}} illustrates this structured pilot pattern in the time domain, while Fig. \ref{sti} illustrates the collected frames for channel estimation. The division of the reception time into $P$ frames is motivated by possible short-term variations caused by the \ac{RIS} imperfections during the channel coherence time. This \textcolor{black}{STI}
model indicates \gil{a more realistic aspect where the RIS elements are subjected to hardware impairments or environmental effects that induces both amplitude and phase perturbations in the RIS elements that are static within each small frame but are non-static from frame-to-frame.} 

By considering an RIS operating under this STI model, the RIS activation pattern is modified in an \textcolor{black}{undesired} 
manner leading to the following impaired reflection pattern in the $k$-th time-block at the $p$-th frame
\begin{equation}
\bar{\mathbf{s}}\left[p,k,t\right] = \left[e_{n,p}\beta_{1,k,t}e^{j\phi_{1,k,t}}, \ldots, e_{N,p}\beta_{N,k,t}e^{j\phi_{N,k,t}}                                       \right]^{\text{T}},
\end{equation}
where $\phi_{n,k,t} \in (0,2\pi]$ and $\beta_{n,k,t} \in \{0,1\}$. Equivalently,
\begin{equation}
\bar{\mathbf{s}}\left[p,k,t\right] = \mathbf{e}\left[p\right] \odot \mathbf{s}\left[k,t\right] \mathbb{C}^{N \times 1},
\label{spk}
\end{equation}
where $\mathbf{s}[k,t] \in \mathbb{C}^{N \times 1}$ is the known activation pattern and $\mathbf{e}\left[p\right] = \left[e_{1,p}, \ldots, e_{N,p}\right]^{\text{T}} \in \mathbb{C}^{N \times 1}$ $\forall p = 1, \ldots, P$, depends on the $p$-th received frame and models the unknown non-static amplitude and phase fluctuations along the training time. The entries of the random vector $\mathbf{e}\left[p\right]$ associated with the $p$-th frame are defined as
\begin{equation}
e_{n,p} = \left\{\begin{array}{ll}
1, & \quad \textcolor{black}{\text{non-impaired case}} \\
\alpha_{n,p}\cdot e^{j\theta_{n,p}}, & \quad \textcolor{black}{\text{otherwise}},
\end{array}
\right.
\label{conds}
\end{equation}
where $0 \leq \alpha_{n,p} \leq 1$ and $0 \leq \theta_{n,p} \leq 2\pi$ $\forall n = 1, \ldots, N$, and $\forall p = 1, \ldots, P$, denote the unwanted amplitude attenuation and phase shift perturbations that \textcolor{black}{affect} the $n$-th \ac{RIS} element at the $p$-th frame, respectively. It is important to note that the model in (\ref{conds}) captures different kinds of real-world \textcolor{black}{imperfections} at the \ac{RIS}. For example, we can note the following situations: 

\begin{itemize}
	\item $\alpha_{n,p} \neq 0$ and $\theta_{n,p} \neq 0$ represent the amplitude absortion and phase shift caused by an object suspended on the $n$-th \ac{RIS} element, or caused by hardware impairments in the electronic circuits that make up the \ac{RIS}. 
	\item $\alpha_{n,p} = 1$ and $\theta_{n,p} \neq 0$ represent the phase noise perturbations from low-resolution phase shifts or phase errors from imperfect channel estimation. 
	\item $\alpha_{n,p} = 0$ represents the maximum absorption i.e., the $n$-th \ac{RIS} element is completely blocked. 
	\item $\alpha_{n,p} = 1$ and $\theta_{n,p} = 0$ represents the \textcolor{black}{non-impaired}
	\ac{RIS} in which no \textcolor{black}{imperfection} affects its $n$-th element. Note that in this ideal case $\bar{\mathbf{s}}\left[p,k,t\right] = \mathbf{s}\left[k,t\right]$ holds since $e_{n,p} = 1$ $\forall n = 1, \ldots, N$, and $\forall p = 1, \ldots, P$.
\end{itemize} 

Therefore, the baseband received pilot signal $\mathbf{y}[p,k,t] \in \mathbb{C}^{L \times 1}$ associated with the $p$-th frame, $k$-th block and $t$-th symbol period can be expressed by
\begin{equation}
    \mathbf{y}[p,k,t] = \mathbf{G}\textrm{diag}(\mathbf{e}\left[p\right] \odot \mathbf{s}\left[k,t\right] )\mathbf{H}^{\textrm{T}}\mathbf{x}[k,t] + \mathbf{v}[p,k,t].
    \label{simples}
\end{equation}

According Fig. \ref{protocol}, we assume that the activation pattern remains constant within the $k$-th block but \textcolor{black}{may vary} between different time-blocks, i,e., $\mathbf{s}[k,t] = \mathbf{s}[k], \,\, \text{for}\,\, t = 1, \dots, T$. Furthermore, the \textcolor{black}{pilot symbol} $\mathbf{x}\left[k,t\right] \in \mathbb{C}^{M \times 1}$ transmitted at the $t$-th symbol period within the $k$-th time-block is reused for each $k = 1, \ldots, K$, i.e., $\mathbf{x}[k,t] = \mathbf{x}[t], \,\, \text{for}\,\, k = 1, \dots, K$. Based on this training protocol, by collecting the received signals during the $T$ symbol periods at the $p$-th frame and $k$-th time-block, the model in (\ref{simples}) can be rewritten as
\begin{equation}
\mathbf{Y}\left[p,k\right] = \mathbf{G}\text{diag}\left(\mathbf{e}\left[p\right] \odot \mathbf{s}\left[k\right]\right)\mathbf{H}^{\text{T}}\mathbf{X} + \mathbf{V}\left[p,k\right],
\label{dec1}
\end{equation}
where $\mathbf{Y}\left[p,k\right] = \left[\mathbf{y}\left[p,k,1\right], \ldots, \mathbf{y}\left[p,k,T\right]\right] \in \mathbb{C}^{L \times T}$ $\forall p = 1, \ldots, P$, and $\forall k = 1, \ldots, K$. The matrices $\mathbf{H} \in \mathbb{C}^{M \times N}$ and $\mathbf{G} \in \mathbb{C}^{L \times N}$ denote the Tx-RIS and RIS-Rx channels, respectively, while $\mathbf{X} = \left[\mathbf{x}\left[1\right], \ldots, \mathbf{x}\left[T\right]\right] \in \mathbb{C}^{M \times T}$ collects the pilot signals transmitted within the $k$-th time-block, and $\mathbf{V}\left[p,k\right] = \left[\mathbf{v}\left[p,k,1\right], \ldots, \mathbf{v}\left[p,k,T\right]\right] \in \mathbb{C}^{L \times T}$ is the additive white Gaussian noise (AWGN) matrix with zero mean and unit variance elements. In order to simplify our formulation and analysis, without loss of \textcolor{black}{generality,} we assume the transmission of the pilot signal $\mathbf{X} = \mathbf{I}_{M}$.

In a more convenient form for our formulation, the received signal in (\ref{dec1}) can be written in its complete matrix and decoupled format as
\begin{equation}
\mathbf{Y}\left[p,k\right] = \mathbf{G}\mathbf{D}_{p}\left(\mathbf{E}\right)\mathbf{D}_{k}\left(\mathbf{S}\right)\mathbf{H}^{\text{T}} + \mathbf{V}\left[p,k\right],
\label{dec2}
\end{equation}
where $\mathbf{S} = \left[\mathbf{s}[1], \dots, \mathbf{s}[K]\right]^{\text{T}} \in \mathbb{C}^{K \times N}$ collects in its rows the known activation pattern used accross $K$ blocks specifically configured for the channel estimation. Each row of $\mathbf{E} = \left[\mathbf{e}\left[1\right], \ldots, \mathbf{e}\left[P\right]\right]^{\text{T}} \in \mathbb{C}^{P \times N}$ collects the unknown amplitude and phase parameters for the \ac{RIS} elements impaired at the $p$-th frame. Throughout this work,
we assume that a number of \textcolor{black}{$N_{B} = N R_{B}$} random elements at the \ac{RIS} are subject to \textcolor{black}{imperfections,} where \textcolor{black}{$R_{B} \in [0,1]$} denotes its occurrence probability.

\section{PARAFAC Modeling}

According to \cite{Kolda}, the noiseless signal part of (\ref{dec2}) expresses the ($p,k$)-th frontal slice of a fourth-order tensor $\mathcal{Y} \in \mathbb{C}^{L \times M \times K \times P}$ that follows the PARAFAC decomposition \textcolor{black}{and} \gil{its representation in terms of $n$-mode product notation is given by}
\begin{equation}
\mathcal{Y} = \mathcal{I}_{4,N} \times_{1} \mathbf{G} \times_{2} \mathbf{H} \times_{3} \mathbf{S} \times_{4} \mathbf{E}.
\label{t4}
\end{equation} 
\textcolor{black}{Here,} $\mathcal{I}_{4,N}$ denotes the fourth-order identity tensor of size $N \times N \times N \times N$, while $\mathbf{G}$, $\mathbf{H}$, $\mathbf{S}$ and $\mathbf{E}$ are the 1,2,3,4-mode factor matrices of the decomposition, respectively.

By stacking column-wise the noiseless received signal in (\ref{dec2}) for the $K$ time-blocks at frame $p$ as the matrix $\mathbf{Y}_{p} = \left[\mathbf{Y}\left[p,1\right], \ldots, \mathbf{Y}\left[p,K\right]\right] \in \mathbb{C}^{L \times MK}$, we have
\begin{equation}
\mathbf{Y}_{p} = \mathbf{G}\mathbf{D}_{p}\left(\mathbf{E}\right)\left[\mathbf{D}_{1}\left(\mathbf{S}\right)\mathbf{H}^{\text{T}}, \ldots,\mathbf{D}_{K}\left(\mathbf{S}\right)\mathbf{H}^{\text{T}}\right], 
\label{f4}
\end{equation}
$\forall p = 1, \ldots, P$. Applying the property (\ref{pp2}) to the right-hand side of (\ref{f4}), a more compact form is obtained as
\begin{equation}
\mathbf{Y}_{p} = \mathbf{G}\mathbf{D}_{p}\left(\mathbf{E}\right)\left(\mathbf{S} \diamond \mathbf{H}\right)^{\text{T}} \in \mathbb{C}^{L \times MK}.
\label{gg}
\end{equation}
From (\ref{gg}), we can define the new column-wise collection $\left[\mathcal{Y}\right]_{(1)} = \left[\mathbf{Y}_{1}, \ldots, \mathbf{Y}_{P}\right] \in \mathbb{C}^{L \times MKP}$ as the 1-mode matrix unfolding of the received signal tensor $\mathcal{Y} \in \mathbb{C}^{L \times M \times K \times P}$ in (\ref{t4}), which is given by
\begin{equation}
\left[\mathcal{Y}\right]_{(1)} = \mathbf{G}\left[\mathbf{D}_{1}\left(\mathbf{E}\right)\left(\mathbf{S} \diamond \mathbf{H}      \right)^{\text{T}}, \ldots, \mathbf{D}_{P}\left(\mathbf{E}\right)\left(\mathbf{S} \diamond \mathbf{H}\right)^{\text{T}}\right].
\label{ll}
\end{equation}
By applying property (\ref{pp2}) to the right-hand side of (\ref{ll}), we finally \textcolor{black}{obtain}
\begin{equation}
\left[\mathcal{Y}\right]_{(1)} = \mathbf{G}\left(\mathbf{E} \diamond \mathbf{S} \diamond \mathbf{H}\right)^{\text{T}} \in \mathbb{C}^{L \times MKP}.
\label{unf41}
\end{equation} 
The remaining 2-mode, 3-mode and 4-mode matrix unfoldings can be deduced using a similar procedure by permuting the factor matrices in (\ref{dec2}). This leads to the following factorizations to the other unfoldings 
\begin{eqnarray}
\left[\mathcal{Y}\right]_{(2)} &=& \mathbf{H}\left(\mathbf{E} \diamond \mathbf{S} \diamond \mathbf{G}\right)^{\text{T}} \in \mathbb{C}^{M \times LKP},\label{unf42} \\
\left[\mathcal{Y}\right]_{(3)} &=& \mathbf{S}\left(\mathbf{E} \diamond \mathbf{H} \diamond \mathbf{G}\right)^{\text{T}} \in \mathbb{C}^{K \times LMP}, \label{hop}\\
\left[\mathcal{Y}\right]_{(4)} &=& \mathbf{E}\left(\mathbf{S} \diamond \mathbf{H} \diamond \mathbf{G}\right)^{\text{T}} \in \mathbb{C}^{P \times LMK}.\label{unf43}
\end{eqnarray}

\section{Proposed HOSVD-STI Algorithm for Joint Channel and Imperfections Estimation}

\textcolor{black}{We now} derive a \textcolor{black}{HOSVD-based} closed-form solution 
for joint estimation of the involved communication channels $\mathbf{G}$ and $\mathbf{H}$ and the RIS imperfections embedded in $\mathbf{E}$. According to (\ref{hop}), the transpose of the 3-mode unfolding of $\mathcal{Y}$ is denoted by
\begin{equation}
\left[\mathcal{Y}\right]^{\text{T}}_{(3)} = \left(\mathbf{E} \diamond \mathbf{H} \diamond \mathbf{G}\right)\mathbf{S}^{\text{T}}.
\label{ho00}
\end{equation}
The first processing step at the receiver is to apply a bilinear time-domain matched-filtering by multiplying both sides in (\ref{ho00}) by the pseudo-inverse of $\mathbf{S}^{\text{T}}$, \textcolor{black}{resulting in}
\begin{equation}
\tilde{\mathbf{Y}} = \mathbf{E} \diamond \mathbf{H} \diamond \mathbf{G} \in \mathbb{C}^{LMP \times N}, 
\label{ho01}
\end{equation} 
where $\tilde{\mathbf{Y}} = \left[\mathcal{Y}\right]^{\text{T}}_{(3)}\left(\mathbf{S}^{\text{T}}\right)^{\dag}$. From (\ref{ho01}), decoupled estimates of the channel matrices and \ac{RIS} \textcolor{black}{imperfections} can be obtained by separating each factor matrix in the Khatri-Rao product. In this sense, the estimates can be obtained by minimizing the following cost function
\begin{equation} 
\underset{\mathbf{G},\mathbf{H},\mathbf{E}}{\text{min}}\left\|\tilde{\mathbf{Y}} - \mathbf{E} \diamond \mathbf{H} \diamond \mathbf{G}\right\|_{\text{F}}^{2}.
\label{optho}
\end{equation}
We propose to solve this problem by means of multiple rank-one tensor approximations via the HOSVD. To this end, let us define $\tilde{\mathbf{Y}} = \left[\tilde{\mathbf{y}}_{1}, \ldots, \tilde{\mathbf{y}}_{N}\right] \in \mathbb{C}^{LMP \times N}$. The $n$-th column of $\tilde{\mathbf{Y}}$ can be written as
\begin{equation}
\tilde{\mathbf{y}}_{n} = \mathbf{e}_{n} \otimes \mathbf{h}_{n} \otimes \mathbf{g}_{n} \in \mathbb{C}^{LMP \times 1},
\label{ho05}
\end{equation} 
where $\mathbf{e}_{n} \in \mathbb{C}^{P \times 1}$, $\mathbf{h}_{n} \in \mathbb{C}^{M \times 1}$ and $\mathbf{g}_{n} \in \mathbb{C}^{L \times 1}$ denote the $n$-th column of $\mathbf{E}$, $\mathbf{H}$ and $\mathbf{G}$, respectively. Using the equivalence property \textcolor{black}{in} (\ref{pp4}) that relates the Kronecker product to the outer product, we can rewrite (\ref{ho05}) as 
\begin{equation}
\tilde{\mathbf{y}}_{n} = \text{vec}\left(\mathbf{g}_{n} \circ \mathbf{h}_{n} \circ \mathbf{e}_{n}\right) \in \mathbb{C}^{LMP \times 1},
\end{equation}
that represents the vectorized form of the following third-order rank-one tensor
\begin{equation}
\tilde{\mathcal{Y}}_{n} = \mathbf{g}_{n} \circ \mathbf{h}_{n} \circ \mathbf{e}_{n} \in \mathbb{C}^{L \times M \times P}.
\label{r1ho}
\end{equation}
Thus, the optimization problem in (\ref{optho}) is equivalent to finding the estimates of $\mathbf{H}$, $\mathbf{G}$ and $\mathbf{E}$ that minimize a set of $N$ rank-one tensor approximations, i.e, 
\begin{equation} 
\left(\hat{\mathbf{G}}, \hat{\mathbf{H}}, \hat{\mathbf{E}}\right) = \underset{\mathbf{G},\mathbf{H},\mathbf{E}}{\text{argmin}}\sum_{n=1}^{N}\left\|\tilde{\mathcal{Y}}_{n} - \mathbf{g}_{n} \circ \mathbf{h}_{n} \circ \mathbf{e}_{n}\right\|_{\text{F}}^{2}.
\label{gj}
\end{equation}

\begin{algorithm}[t]
	\caption{\small{HOSVD-STI Algorithm}}
	\begin{algorithmic}\label{Alg3}
		\STATE \small{\textbf{for} $n = 1, \ldots, N$
			\STATE \textbf{1.} \textit{Rearrange the $n$-th column of $\tilde{\mathbf{Y}}$ in Equation (\ref{ho01})} \\
			\STATE \quad \textit{as the rank-one tensor $\tilde{\mathcal{Y}}_{n}$ in Equation (\ref{r1ho})};  
			\STATE \textbf{2.} \textbf{HOSVD procedure}
			\STATE \quad \textbf{2.1} \textit{Compute} $\mathbf{U}_{n}^{(1)}$ \textit{as the} $L$ \textit{left singular vectors of} $\left[\tilde{\mathcal{Y}}_{n}\right]_{(1)}$:   
			\begin{displaymath}
			\left[\tilde{\mathcal{Y}}_{n}\right]_{(1)} = \mathbf{U}_{n}^{(1)} \cdot \mathbf{\Sigma}_{n}^{(1)} \cdot \mathbf{V}_{n}^{(1)\text{H}};
			\end{displaymath}
			\STATE \quad \textbf{2.2} \textit{Compute} $\mathbf{U}_{n}^{(2)}$ \textit{as the} $M$ \textit{left singular vectors of} $\left[\tilde{\mathcal{Y}}_{n}\right]_{(2)}$:   
			\begin{displaymath}
			\left[\tilde{\mathcal{Y}}_{n}\right]_{(2)} = \mathbf{U}_{n}^{(2)} \cdot \mathbf{\Sigma}_{n}^{(2)} \cdot \mathbf{V}_{n}^{(2)\text{H}};
			\end{displaymath}
			\STATE \quad \textbf{2.3} \textit{Compute} $\mathbf{U}_{n}^{(3)}$ \textit{as the} $P$ \textit{left singular vectors of} $\left[\tilde{\mathcal{Y}}_{n}\right]_{(3)}$:   
			\begin{displaymath}
			\left[\tilde{\mathcal{Y}}_{n}\right]_{(3)} = \mathbf{U}_{n}^{(3)} \cdot \mathbf{\Sigma}_{n}^{(3)} \cdot \mathbf{V}_{n}^{(3)\text{H}};
			\end{displaymath}
			\STATE \quad \textbf{2.4} \textit{Compute the HOSVD core tensor} $\mathcal{G}_{n}$ \textit{as}: 
			\begin{displaymath}
			\mathcal{G}_{n} = \tilde{\mathcal{Y}}_{n} \times_{1} \mathbf{U}_{n}^{(1)\text{H}} \times_{2} \mathbf{U}_{n}^{(2)\text{H}} \times_{3} \mathbf{U}_{n}^{(3)\text{H}};
			\end{displaymath}
			\STATE \quad \textbf{end procedure}
			\STATE \textbf{3.} \textit{Obtain the estimates for} $\hat{\mathbf{g}}_{n}$, $\hat{\mathbf{h}}_{n}$ and $\hat{\mathbf{e}}_{n}$ \textit{from Equations} \\
			\STATE \quad (\ref{zs1}), (\ref{zs2}) \textit{and} (\ref{zs3}), \textit{respectively;}  
			\STATE \textbf{end}
			\STATE \textbf{4.} \textit{Return the matrices} $\hat{\mathbf{G}} = \left[\hat{\mathbf{g}}_{1}, \ldots, \hat{\mathbf{g}}_{N}\right]$, $\hat{\mathbf{H}} = \left[\hat{\mathbf{h}}_{1}, \ldots, \hat{\mathbf{h}}_{N}\right]$ \\
			\STATE \quad \textit{and} $\hat{\mathbf{E}} = \left[\hat{\mathbf{e}}_{1}, \ldots, \hat{\mathbf{e}}_{N}\right]$.
		}
	\end{algorithmic}
\end{algorithm}

Let us introduce the HOSVD of $\tilde{\mathcal{Y}}_{n}$ as 
\begin{equation}
\tilde{\mathcal{Y}}_{n} = \mathcal{G}_{n} \times_{1} \mathbf{U}_{n}^{(1)} \times_{2} \mathbf{U}_{n}^{(2)} \times_{3} \mathbf{U}_{n}^{(3)} \in \mathbb{C}^{L \times M \times P},
\end{equation}
where $\mathbf{U}_{n}^{(1)} \in \mathbb{C}^{L \times L}$, $\mathbf{U}_{n}^{(2)} \in \mathbb{C}^{M \times M}$ and $\mathbf{U}_{n}^{(3)} \in \mathbb{C}^{P \times P}$ are unitary matrices, while $\mathcal{G}_{n} \in \mathbb{C}^{L \times M \times P}$ denotes the HOSVD core tensor.
The estimates of the vectors $\mathbf{g}_{n}$, $\mathbf{h}_{n}$ and $\mathbf{e}_{n}$ that \textcolor{black}{solve} the LS problem in (\ref{gj}) can be obtained by truncating the HOSVD of $\tilde{\mathcal{Y}}_{n}$ to its dominant rank-one component, \textcolor{black}{yielding}    
\begin{eqnarray}
\hat{\mathbf{g}}_{n} &=& \sqrt[3]{\left(\mathcal{G}_{n}\right)_{1,1,1}} \cdot \mathbf{u}^{(1)}_{1,n}, \label{zs1}\\
\hat{\mathbf{h}}_{n} &=& \sqrt[3]{\left(\mathcal{G}_{n}\right)_{1,1,1}} \cdot \mathbf{u}^{(2)}_{1,n}, \label{zs2}\\
\hat{\mathbf{e}}_{n} &=& \sqrt[3]{\left(\mathcal{G}_{n}\right)_{1,1,1}} \cdot \mathbf{u}^{(3)}_{1,n}, \label{zs3}
\end{eqnarray} 
where $\mathbf{u}^{(1)}_{1,n} \in \mathbb{C}^{L \times 1}$, $\mathbf{u}^{(2)}_{1,n} \in \mathbb{C}^{M \times 1}$ and $\mathbf{u}^{(3)}_{1,n} \in \mathbb{C}^{P \times 1}$ are the first higher order singular vectors, i.e., the first column of $\mathbf{U}_{n}^{(1)}$, $\mathbf{U}_{n}^{(2)}$ and $\mathbf{U}_{n}^{(3)}$, respectively. \textcolor{black}{Here,} $\left(\mathcal{G}_{n}\right)_{1,1,1}$ is the first element of the core tensor $\mathcal{G}_{n}$. The estimates of $\hat{\mathbf{G}}$, $\hat{\mathbf{H}}$ and $\hat{\mathbf{E}}$ are obtained by repeating the procedure \textcolor{black}{of (\ref{r1ho})-(\ref{zs3})} for the $N$ columns of $\tilde{\mathbf{Y}}$ in (\ref{ho01}). In other words, a total of $N$ rank-one tensor approximations via HOSVD are necessary to obtain the full estimates of the matrices $\hat{\mathbf{G}} = \left[\hat{\mathbf{g}}_{1}, \ldots, \hat{\mathbf{g}}_{N}\right]$, $\hat{\mathbf{H}} = \left[\hat{\mathbf{h}}_{1}, \ldots, \hat{\mathbf{h}}_{N}\right]$ and $\hat{\mathbf{E}} = \left[\hat{\mathbf{e}}_{1}, \ldots, \hat{\mathbf{e}}_{N}\right]$ in a closed-form manner. The implementation steps of the proposed closed-form HOSVD-STI algorithm are summarized in Algorithm 1. 

\subsection{Indentifiability and Computational Complexity}

The proposed HOSVD-STI algorithm is a closed-form solution requiring only that the \ac{RIS} activation pattern matrix $\mathbf{S} \in \mathbb{C}^{K \times N}$ have full column-rank in order to guarantee the uniqueness in the LS sense when the bilinear time-domain matched-filtering preprocessing is performed at the receiver side, as indicated in (\ref{ho01}). This leads to $K \geq N$ as a necessary and sufficient condition. 
It can be satisfied under the truncated discrete Fourier transform (DFT) design for the RIS activation pattern matrix since it has orthonormal columns, \gil{i.e., $\mathbf{S}^{\textrm{S}}\mathbf{H} = K\mathbf{I}_{N}$}. Therefore, by satisfying $K \geq N$, the estimated matrices $\hat{\mathbf{G}}$, $\hat{\mathbf{H}}$ and $\hat{\mathbf{E}}$ are column scaled versions of their true values. This scaling ambiguity can be eliminated with a simple normalization procedure. 

\textcolor{black}{Regarding} the computational complexity of the HOSVD-STI algorithm, it is dominated by the HOSVD computation of the third-order rank-one tensor in (\ref{r1ho}), which is equivalent to compute the truncated singular value decompositions (SVDs) of its 1-mode, 2-mode and 3-mode unfolding matrices to rank-one. These truncated SVDs are repeated $N$ times. Therefore, the HOSVD-STI algorithm has a complexity $\mathcal{O}\left(NMLP\right)$ flops. However, the execution of the HOSVD-STI algorithm can be parallelized if more than one processor is available leading to greater reduction in the total processing time for channel estimation, making it attractive when low processing delay is needed. 

\section{Simulation Results}
\begin{figure}[!t]
\centering
\subfloat[\textcolor{black}{NMSE of $\hat{\mathbf{H}}$ \textit{versus} SNR (dB).}]{\includegraphics[scale=0.55]{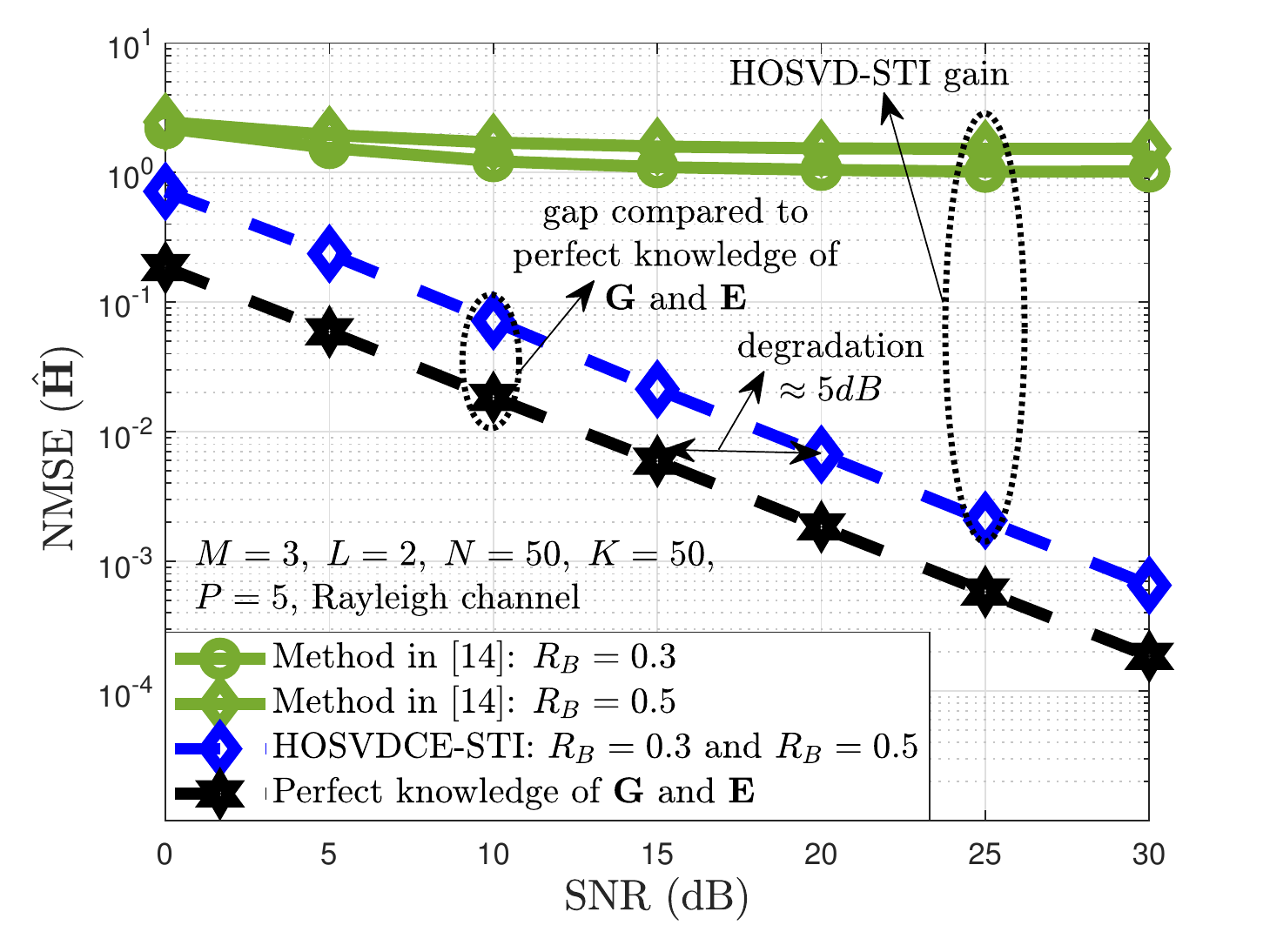}\label{iti1}}\\
\subfloat[\textcolor{black}{NMSE of $\hat{\mathbf{G}}$ \textit{versus} SNR (dB).}]{\includegraphics[scale=0.55]{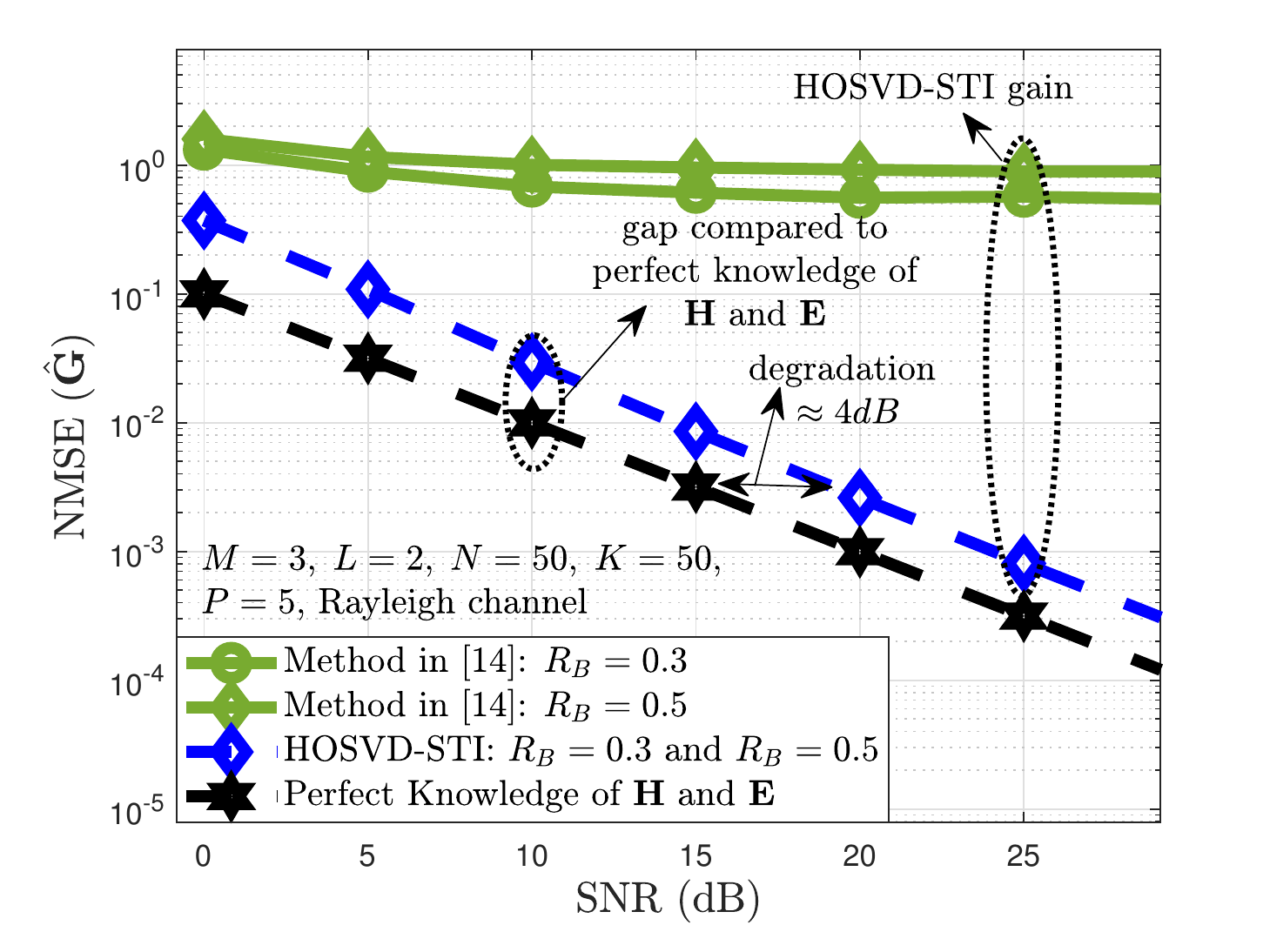}\label{iti2}}\\
\subfloat[\textcolor{black}{NMSE of $\hat{\mathbf{E}}$ \textit{versus} SNR (dB).}]{\includegraphics[scale=0.55]{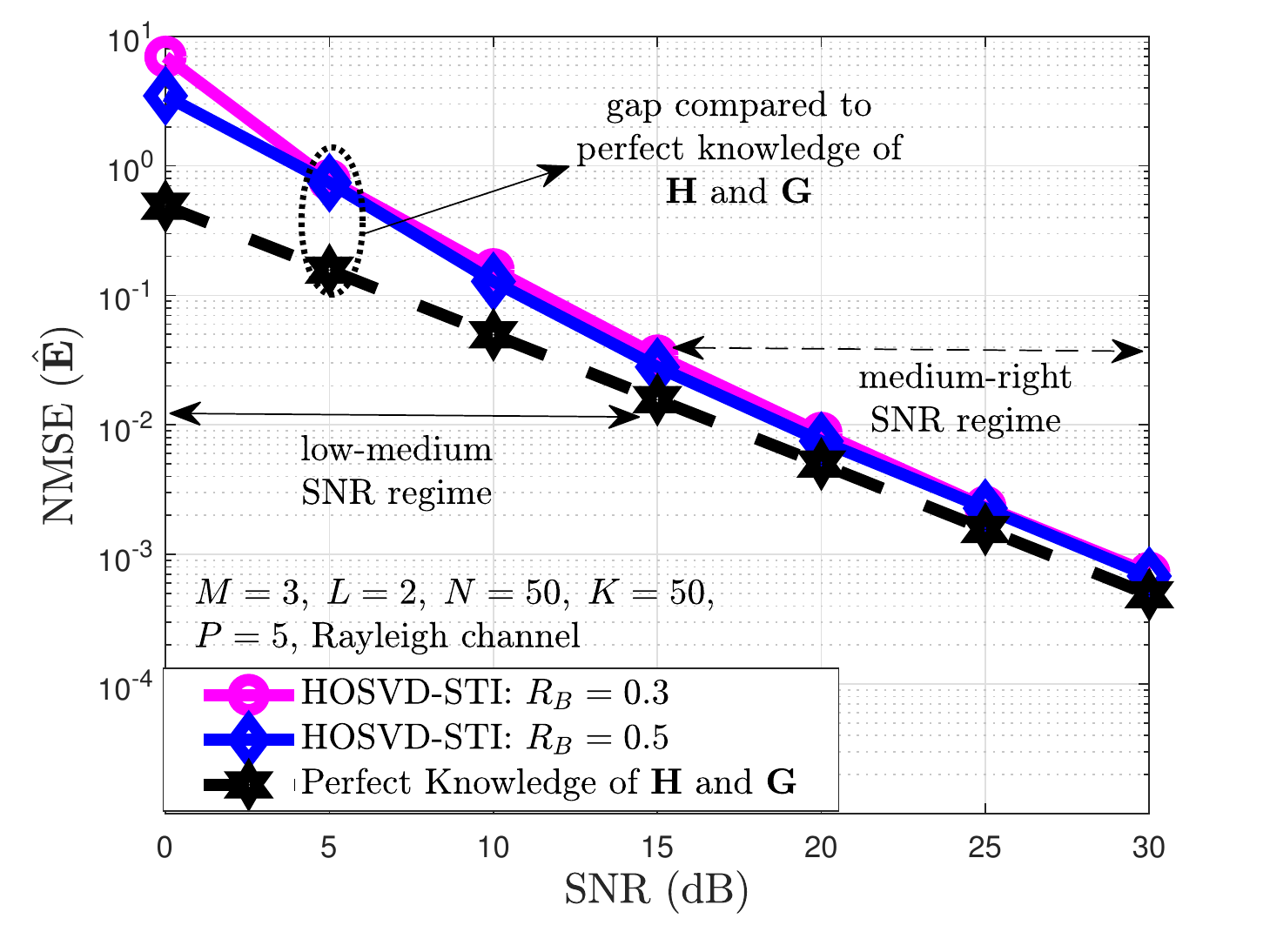}\label{iti3}}
\caption{\textcolor{black}{NMSE of HOSVD-STI algorithm \textit{versus} SNR (dB) \textcolor{black}{assuming i.i.d. Rayleigh channels}.}}
\end{figure}

We present a set of simulation results for performance evaluation of the proposed tensor-based HOSVD-STI algorithm. The results presented here are averaged over \textcolor{black}{$\Omega = 3000$} independent Monte Carlo runs. Each run corresponds \textcolor{black}{to} a different realization of the Tx-RIS and RIS-Rx channels, \ac{RIS} patterns, impairment parameters and noise. We have designed the \ac{RIS} pattern matrix $\mathbf{S}$ as a DFT matrix. The amplitude and phase impairments parameters embedded in the matrix $\mathbf{E}$ \textcolor{black}{follow a} uniform distribution between 0 and 1 and between 0 and $2\pi$, respectively. The location of the $N_{B}$ impaired elements are assumed to be random with occurrence probability $R_{B}$, totalizing $N_{B} = NR_{B}$ impaired elements at the RIS. The metric used to evaluate the estimation accuracy is the \ac{NMSE} between the true and estimated matrices that provides a relative measure for the estimation error of the proposed algorithm. For the estimated channel $\hat{\mathbf{H}}$ we define $\text{NMSE}(\hat{\mathbf{H}}) = \frac{1}{\Omega}\sum_{\omega=1}^{\Omega}\frac{\|\mathbf{H}^{(\omega)} - \hat{\mathbf{H}}^{(\omega)}\|_{\text{F}}^{2}}{\|\mathbf{H}^{(\omega)}\|_{\text{F}}^{2}},$
where \textcolor{black}{$\mathbf{H}^{(\omega)}$} and \textcolor{black}{$\hat{\mathbf{H}}^{(\omega)}$} denote the true channel and its estimate both related to the \textcolor{black}{$\omega$-th run,} respectively. Similar definitions apply to the estimates of $\hat{\mathbf{G}}$ and $\hat{\mathbf{E}}$.

We examine, in Figs. \ref{iti1}, \ref{iti2} and \ref{iti3}, the estimation performance of the proposed HOSVD-STI algorithm. We compare our approach with the iterative bilinear alternating least squares based algorithm proposed in \cite{GilJournal}. It is a tensor-based pilot-assisted method formulated to the ideal case in which no impairments affect the RIS. Additionaly, as a lower-bound, we also plot the \textit{clairvoyant} estimator that assumes perfect knowledge of the channels and impairments. By considering different blockage probabilities, we can observe that the method in \cite{GilJournal} is not suitable to tackle the channel estimation problem when impairments are present. In contrast, \textcolor{black}{our} proposed HOSVD-STI algorithm \textcolor{black}{provides} accurate channel estimates that are not sensitive to the number of impaired elements and closely of the lower-bound, confirming the effectiveness of the proposed algorithm. However, in terms of imperfections estimation, we can see that the HOSVD is sensitive to SNR values providing accurate estimates very close to the lower-bound in the medium-to-right SNR regime, while some performance degradation is observed in the low-to-medium SNR regime.

\begin{figure}[!t]
	\centering
	\includegraphics[scale=0.55]{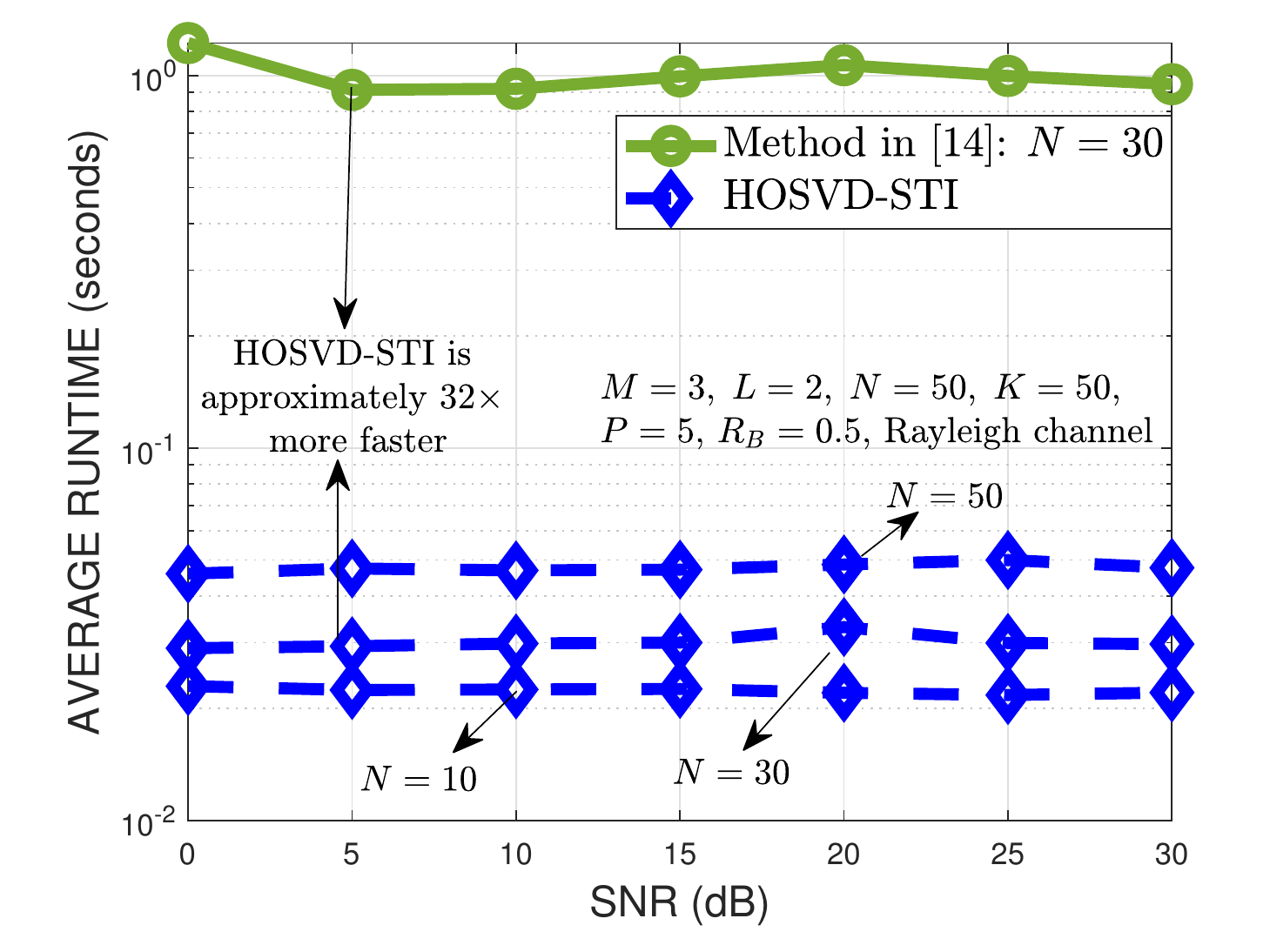}
	\caption{Average runtime (in seconds) of HOSVD-STI algorithm \textit{versus} SNR (dB) assuming i.i.d. Rayleigh channels.}
	\label{mpt}
\end{figure}

In Fig. \ref{mpt}, we evaluate the performance of the HOSVD-STI algorithm in terms of runtime (in seconds). It can be seen that the runtime grows when the number of RIS elements increases, which is an expected result. However, the runtime required by HOSVD-STI algorithm is not sensitive to the SNR since it is a closed-form approach. Note that the closed-form HOSVD-STI algorithm requires less runtime in all the SNR regime, \textcolor{black}{compared to \cite{GilJournal}.} On the other hand, the iterative method in \cite{GilJournal} has a higher computational complexity compared to our proposed algorithm, providing to be less attractive than our approach in both estimation accuracy and complexity performance.

\section{Conclusion}

We have proposed an efficient and low-complexity tensor-based algorithm for joint channel and imperfections estimation in impaired RIS-assisted networks. By assuming the STI model in which the behavior of the RIS impairments are non-static with respect to the channel coherence time, and resorting to the inherent tensor algebraic structure of the received signal, \textcolor{black}{we have formulated} the HOSVD-STI algorithm. It is a closed-form solution that presents lower computational complexity and parallel processing capability compared to the competing tensor-based iterative alternating solution. \textcolor{black}{Our proposed HOSVD-STI algorithm} proves to be robust when operating in more realistic scenarios in which the RIS elements are subject to hardware impairments and/or environmental effects providing accurate channel estimates in this non-idealized case.


\section*{Acknowledgment}
This work was supported in part by the Ericsson Research, Technical Cooperation UFC.48, by the Coordenação de Aperfeiçoamento de Pessoal de Nível Superior - Brasil (CAPES) and by CNPq. \textcolor{black}{Gábor Fodor was partially supported by the Digital Futures project PERCy.}



\end{document}